\documentstyle{l-aa}

\newcommand{\be}{\begin{equation}} 
\newcommand{\ee}{\end{equation}}

\def\lsim{\lower.5ex\hbox{$\; \buildrel < \over \sim \;$}}
\def\gsim{\lower.5ex\hbox{$\; \buildrel > \over \sim \;$}} 

\def\pmb#1{\setbox0=\hbox{#1}%
  \kern-0.015em\copy0\kern-\wd0
  \kern0.03em\copy0\kern-\wd0
  \kern-0.015em\raise0.0433em\box0 }

\begin{document}
\thesaurus{02 (02.01.2) 08 (08.02.3 LMXBs) 13 (13.25.1; 13.25.3 QPOs)}
\title{Rotational splitting effect in neutron star QPOs}
\author{L. Titarchuk \inst{1,2} and A. Muslimov \inst{1}}
\offprints{L.~Titarchuk}
\institute{1 Laboratory for High Energy Astrophysics, Goddard Space 
              Flight Center Greenbelt MD 20771 USA\\
           2 George Mason University/CSI Fairfax VA 22030 USA}
\date{Received;accepted}
\maketitle
\begin{abstract}
We explain the peak spacing in the power spectra of 
millisecond variability detected by the Rossi X-ray
Timing Explorer (RXTE) in the X-ray emission from the LMXBs 
(Sco X-1, 4U1728-34, 4U1608-522, 4U1636-536, etc) in
terms of the rotational splitting of an intrinsic 
frequency (which is of order of the local Keplerian frequency) caused 
by an accretion disk. We calculate this effect and demonstrate 
that there is a striking agreement with the observational data. 
We show that the observed discrete frequencies ranging from 200 to
1200 Hz can be described by a whole set of overtones. For higher 
overtones (lower frequencies, $\lsim $ 100 Hz) the ratio between 
frequencies is determined by the quantum numbers alone. We suggest 
that a similar phenomena should be observed in 
Black Hole (BH) systems for which the QPO (quasi-periodic oscillation)
frequency should be inversely proportional to the mass of the 
compact object. 
For BH systems the characteristic frequency of oscillations should 
therefore be a factor of 5-10 less than for a neutron star system. 
\end{abstract}
\section{Introduction}

The launch of RXTE opened up a new era in the study of quasi-periodic 
oscillations (QPOs). Recently, kHz QPOs have been discovered in the 
persistent fluxes of eight LMXBs: Sco X-1 (van der Klis et al. 1996a), 
4U1728-43 (Strohmayer et al. 1996a), 4U 1608-52 (Berger et al. 1996), 
4U 1636-53 (Zhang et al. 1996), 4U 0614+091 (Ford et al. 1996), 
4U 1735-44 (Wijnands et al. 1996), 4U 1820-30 (Smale et al. 1996), 
and GX5-1 (van der Klis et al. 1996).
In addition, episodic and nearly coherent oscillations have been 
discovered during several type-I X-ray bursts from KS 1731-260 with a 
frequency of 363 Hz (Strohmayer et al. 1996a), during one type-I burst from 
KS 1731-260 with a frequency of 524 Hz (Morgan \& Smith 1996). In 
addition during 3 bursts from vicinity of GRO J1744-28, oscillations 
with a frequency of 589
Hz (Strohmayer et al. 1996a), and during 4 type-I bursts from 4U 
1636-53 with a frequency of 581 Hz (Zhang et al. 1996) were seen. 
The presence of a
pair of kHz QPOs is characteristic of almost all these observations.
The centroid frequency of these QPOs ranges from 400 (4U 0614+091, 
Ford et al. 1996) to 1171 Hz (4U 1636-53, van der Klis et al. 1996c). 
Also, it is worthwhile to point out that van der Klis et al. (1997) 
reported that they detected, simultaneously with the kHz QPO, two 
additional peaks near 40 and 90 Hz. 
 
In this {\it Letter} we shall demonstrate that the effect of a rotational
splitting (caused by the Keplerian accretion disk) of oscillation frequencies 
that has been ignored in all previous theoretical studies of QPOs 
may be important for our understanding of a phenomenon of QPOs 
in binary systems with neutron stars and black holes. Our conclusion 
about relevance of this effect to the oscillations in the innermost 
region of accretion disk around a compact object is robust and 
does not depend on a particular model of a disk. If the QPO 
phenomenon is associated with such oscillations, then we must observe the 
rotationally split frequencies. The detailed study of this effect 
may provide us with very important information about the physics 
of accretion disks.  

\section{The Effect of Rotational Splitting by an Accretion Disk}

It is well known (see e.g. \cite{unn79}, and references 
therein) that in a rotating star the non-radial
oscillations are split: $\rm \sigma _{klm} = \sigma _{kl} + m 
\Omega _{\ast } C_{kl}$, where $\rm \sigma _{kl}$ is the frequency 
in the nonrotating star, $\rm \sigma _{klm}$ is the frequency 
in a rotating star (in the corotating frame), m is an azimuthal 
number, $\Omega _{\ast }$ is 
the stellar rotation frequency, and $\rm C_{kl}$ is expressed in terms
of an integral that depends on the
stellar structure and on the eigenfunction. Thus, in a rotating star,
a nonradial oscillation drifts at a rate $\rm m \Omega _{\ast }C_{kl}$
relative to a fixed longitude of the star. The effect of rotational 
splitting is due to the Coriolis force and is similar to the 
Zeeman effect in a magnetic field. 

The oscillations of a rotating disk must split as well. The 
rotational splitting of the disk oscillations is particularly 
important for those modes whose frequencies are comparable to 
the local rotation frequency of a disk relative to a distant observer. 
The detailed theory for a star is given 
by Ledoux \& Walraven (1958) and by  Unno et al. (1979).  
Here we present calculations of the splitting effect for an
oscillation frequency due to the Coriolis force in a disk geometry.

The general formula for the correction to the oscillation frequency 
of a star/disk produced by the Coriolis force reads (cf. \cite{unn79},
equation [18.28])

\be
\rm \sigma^{(1)}={{-i \int_{V}[{\pmb{$\Omega $}}
\times {\pmb{$\xi $}}^{(0)}_{k,m}] {\pmb{$\xi $}}^{(0) \ast}_{k,m} dV}
\over{\int_{V}{\pmb{$\xi $}}^{(0)}_{k,m}{\pmb{$\xi $}}^{(0) 
\ast}_{k,m} dV}} .
\ee
Here $\pmb{$\Omega $}$ is the local angular velocity in a disk in the
laboratory frame of reference, $\rm {\pmb{$\xi $}}^{(0)}_
{(k,m)}(t,r,\varphi,z)$ is the (k,m)-displacement component, 
and the integration is over the volume of a disk/ring-like configuration.

Let us consider the oscillations localized in some annulus of a disk
at radius R. We assume that the amplitude of these oscillations decreases 
toward the innermost edge of a disk. Then, the component of a 
displacement vector can be calculated by using the $\rm
(k,m)$-harmonics of a complete set of eigenfunctions for a 
disk/ring-like configuration 
$\rm \{U_{k,m}=r u_{k,m}(r,\varphi ,z,t)\}$, where $\rm {r,\varphi
,z}$ are cylindrical coordinates, and t is the time in a 
laboratory frame of reference. In equation (1) the integration has
to be taken over the volume of a ring of radius R, width $\rm \Delta
R$, and half-thickness H. The vectors $\rm {\pmb{$\xi $}}^{(0)}_
{(k,m)}(t,r,\varphi ,z)$ are usually presented in such a form where 
their components are proportional to the corresponding components of a
gradient of a potential $\rm {U_{k,m}=r u_{k,m}}$. In cylindrical 
coordinates they can be written as (cf. Unno et al. 1979, equation [18.28])
\be
\rm {\pmb{$\xi $}}^{(0)}_{k,m}=\left[\xi_r, \xi_{\varphi} 
{{\partial}\over{\partial \varphi}}, \xi_z r{{\partial}
\over{\partial z}} \right] u_{km} .
\ee
The functions $\rm u_{k,m}$ satisfy the free-boundary conditions 
at $\rm z=\pm H$ (see e.g. Morse
\& Feshbach 1953): 
\be
\rm {{\partial u_{k,m}}\over {\partial z}}(t,H, \varphi)
={{\partial u_{k,m}}\over {\partial z}}(t,-H,\varphi)=0 .
\ee
Also, these functions must be symmetric about the disk plane
\be
\rm u_{k,m}(t,z,\varphi )=u_{k,m} (t,-z,\varphi ) ,
\ee
where $\rm 0\leq z\leq H$.

In the case of uniform rotation, the temporal and azimuthal dependences 
of eigenfunctions $\rm u_{k,m}$ are given by 
$\rm \exp[i(m\varphi-\Omega_0 t)]$, where $\Omega _0$ is the angular 
eigenfrequency of the oscillation (see Unno et al. 1979),  
thus, the component $\rm u_{k,m}$ can be written as 
\be
\rm u_{k,m}=e^{im\varphi}\cos(\pi kz/H)e^{-i\Omega_0 t} ,
\ee
where $\rm m=\pm 1,\pm 2,....$, and $k=1,2,...$ .
  
Note that the functions $\rm u_{k,m}$ are eigenfunctions 
of a two-dimensional Laplace operator. They satisfy the free-boundary 
and symmetry conditions over a z-coordinate (see equations [3],[4]) 
and the periodicity condition over a $\varphi $-coordinate. It is
important that these eigenfunctions form a complete set of functions 
(a fundamental system) which can be used as a basis for the expansion 
(see e.g. Morse \& Feshbach 1953) of any arbitrary displacement 
$\rm {\pmb{$\xi $}}^{(0)}$.  

The numerator and denominator in formula (1) are given by
\be
\rm I_1=-2\pi H R\Delta R(2\Omega m \hat\xi_r\hat \xi_{\varphi}) ,
\ee
and 
\be
\rm I_2=2\pi H R\Delta R[\hat \xi_r^2+(\hat \xi_{z} R/H)^2\pi^2k^2+ 
m^2\xi_{\varphi}^2] ,
\ee
respectively.
Here $\rm \hat \xi_r$, $\hat \xi_{\varphi}$, and $\rm \hat \xi_{z}$ are
the average-weighted displacements for a disk/ring-like configuration 
in radial, azimuthal, and vertical directions, respectively. 
To illustrate the effect of rotational splitting for a disk/ring-like 
configuration we may justifiably assume that 
$\rm \hat\xi_r\sim\hat \xi_{\varphi}$, then we arrive at
\be
\rm \sigma^{(1)}=- {{2m\Omega}\over{s\pi^2k^2+m^2+1}} ,
\ee
where $\rm s(R/H)=(\hat\xi_{z}R/\hat\xi_{r}H)^2$ is a function
determined by the vertical structure of a disk. For more or less 
realistic structure of a disk $\rm s\lsim 1$ and depends on the ratio 
$\rm R/H$, which is either $\gsim 1$ (near a compact object) or 
$\gg 1$ (far away from a compact object). We must note that the 
relationship between $\rm \hat \xi_r$, $\hat \xi_{\varphi}$, 
and $\rm \hat \xi_{z}$ depends also on the azimuthal and vertical 
mode numbers. For example, consider an individual oscillation mode 
with displacement $\rm {\pmb{$\xi $}}^{(0)}_{(k,m)}$ in an 
incompressible
fluid. Then $\rm \nabla \cdot {\pmb{$\xi $}}^{(0)}_{(k,m)}=0$, and we
get a linear partial differential equation that determines 
a relationship between $\rm \hat \xi_r$, $\rm \hat \xi_{\varphi}$, and
$\rm \hat \xi_{z}$. This relationship, according to expression (2), 
should depend on the 
azimuthal and vertical mode numbers, m and k, respectively.

Thus, the frequency of the oscillations detected by a distant observer 
is given by (cf. \cite{unn79}, equation [18.33])
\begin{eqnarray}
\rm \Omega_{k,m} & \equiv & \rm \Omega_0+m\Omega+\sigma^{(1)} =
\nonumber \\ 
& & \rm = \Omega_0+m \Omega \left(1-{2 \over {s\pi^2k^2+m^2+1}} 
\right) ,
\end{eqnarray}
where m and k are the azimuthal and vertical mode numbers,
respectively.

\section{Explanation of QPO Frequencies in Terms of Rotational Splitting}

To estimate the frequency splits, we may adopt 
$\Omega_0=\Omega $, which is a very good approximation for 
a nearly Keplerian accretion
disk. In this case, for the lowest-order modes with $\rm m = 0,~-1$ and 
k = 1, 2, 3, 4, and 5 the oscillation frequencies can be calculated as 
$$
\rm \Omega_{k,0} =  \Omega_0 ,
$$
$$
\rm \Omega_{1,-1} = 2\Omega_0 /(s\pi^2+2) ,
$$
$$
\rm \Omega_{2,-1} = 2\Omega_0 /(s4\pi^2+2) ,
$$
$$
\rm \Omega_{3,-1} = 2\Omega_0 /(s9\pi^2+2) ,
$$
$$
\rm \Omega_{4,-1} = 2\Omega_0 /(s16\pi^2+2) , 
$$
and 
$$
\rm \Omega_{5,-1} =2\Omega_0 /(s25\pi^2+2) ,
$$ 
respectively. 

\noindent
For $\rm m=-2$ and $\rm k=1$ we have 
$$
\rm |\Omega_{1,-2}|=\Omega_0 [1-4/(s\pi^2+5)].
$$

Let us assume that $\rm \nu _0 \equiv \Omega _0/2\pi 
= 1200$ Hz. Then, for the most plausible range of $\rm s \approx 0.7-1$
we get the following frequencies (arranged in ascending order, in Hz): 
$$
\rm \nu _{5,-1} \approx 10-14~~,
$$
$$
\rm \nu _{4,-1} \approx 15-20~~,
$$
$$
\rm \nu _{3,-1} \approx 30-40~~,
$$
$$
\rm \nu _{2,-1} \approx 60-80~~,
$$
$$
\rm \nu _{1,-1} \approx 200-300~~,
$$
and 
$$
\rm \nu _{1,-2} \approx 800-880~~.
$$ 
These frequencies perfectly match the observed QPO frequencies in
LMXBs. It is important that for the modes with a
low azimuthal number $\rm m = -1$ and high k-numbers (high overtones 
in the vertical direction) the ratio between frequencies does not depend on 
the function s, rather it is solely determined by a quantum number k: 
$\rm \nu _{k,-1}/\nu _{k+1,-1} \approx (k+1)^2/k^2$  ($\rm k\geq 2$).  
The frequencies of modes with $\rm m = -2$ and $\rm k\geq 2$
concentrate to $\rm \nu _{k,-2} \approx \nu _0$. It must be pointed
out that the modes with $\rm m = 1, 2, 3, ...$ and $\rm k \geq 1$ 
have frequencies concentrating to the frequency $\rm (m+1) \nu _0$. 
Perhaps, these modes are not allowed or, for some reason, are not 
excited. In this regard, it would be interesting to investigate 
the selection rule for the oscillations in the innermost part of 
accretion disk.

It is important to note that a contribution of each individual 
$\rm (k,m)-$component to the power spectrum of oscillations is 
determined by the smoothness of a function characterizing the
perturbed surface. It is well-known from the Fourier analysis
that the power of the {\it l}-component is $\rm \sim {\it l}^{-2N}$,
where N is the order of highest existing derivative of
a function describing the perturbed surface with respect to 
the corresponding coordinate. In the case of the oscillations of 
the innermost part of accretion disk, $\rm {\it l} = m$ or 
$\rm {\it l} = k$ for the azimuthal or vertical perturbations 
respectively. It is very likely that the azimuthal perturbations 
are smooth, so that we may expect
the presence of only the lowest order modes with $\rm |m|=0,1,2
$ in the power spectrum. On the contrary, the vertical perturbations 
are essentially discontinuous, and their Fourier spectra should contain a
large number of modes with $\rm k > 1$, with the contribution of each
of these modes to the power spectrum $\rm \sim k^{-2}$.

The occurrence of the localized oscillations in the disk may indicate 
that there is a local release of (e.g. gravitational) energy 
in addition to the energy of viscous dissipation. If so, then the
local plasma temperature should be $\rm T \sim 10-20$ keV, which is much 
higher than the temperature in a standard viscous disk (see Shakura \&
Sunyaev 1973). The possibility of such oscillations,
their energetics, high quality Q, and substantial modulation of the 
observed X-ray luminosity due to these oscillations, are discussed 
by Titarchuk et al. (1997) in the context of the effect of a 
centrifugal barrier which is thought to take place in a 
boundary region surrounding a neutron star.

\section{Summary}
        
We suggest that in the QPO phenomenon we observe the effect of 
a rotational splitting of the oscillation frequency. For
a Keplerian disk this effect should be most pronounced because the  
characteristic frequency of the oscillations of 
a centrigugal-barrier region is of the order of the Keplerian
frequency.
 
We conclude that the observed three frequencies ranging from 300 to 1170 Hz 
can be naturally interpreted in terms of a rotational splitting of 
the main frequency: the lowest frequency, $\sim $ 200-300 Hz, corresponds to 
a mode with $\rm m = - 1$ and $\rm k = 1$; the higher 
frequency, $\sim 800$ Hz, corresponds to a mode with $\rm m = - 2$ and 
$\rm k = 1$, and the highest frequency is the main frequency of
the oscillations. The issue of the excitation of these modes and emission 
mechanism will be addressed elsewhere.  

\acknowledgements{L.T. would like to acknowledge support from NASA grant NAG
5-3240, and A.M. thanks NASA and NRC Senior Research Associateship 
at the LHEA in GSFC. The authors also thank Jean Swank and Chris Shrader 
for discussion and Ralph Wijers for a thorough reading to improve 
the final version of the manuscript.}

\end{document}